\documentclass[conference]{IEEEtran}
\usepackage{flushend}
\usepackage{cite}
\usepackage{amsmath,amssymb,amsfonts}
\usepackage{algorithmic}
\usepackage{graphicx}
\usepackage{textcomp}
\usepackage{xcolor}

\usepackage[utf8]{inputenc}

\usepackage{url}
\usepackage{balance}
\usepackage{array}
\usepackage{verbatim} 
\usepackage{booktabs,ragged2e}
\usepackage[flushleft]{threeparttable}
\usepackage{siunitx}
\usepackage[capitalise]{cleveref}
\usepackage[version=4]{mhchem} 
\usepackage{verbatim}
\usepackage{makecell}

\def\BibTeX{{\rm B\kern-.05em{\sc i\kern-.025em b}\kern-.08em
    T\kern-.1667em\lower.7ex\hbox{E}\kern-.125emX}}
\begin{document}

\title{In-Sensor Motion Recognition with Memristive System and Light Sensing Surfaces\\
}

\author{\IEEEauthorblockN{ Hritom Das, Imran Fahad, SNB Tushar, Sk Hasibul Alam, \\ Graham Buchanan, Danny Scott, Garrett S. Rose, and Sai Swaminathan}
 \IEEEauthorblockA{Min H. Kao Department of Electrical Engineering \& Computer Science\\
 The University of Tennessee, Knoxville\\
 Knoxville, TN 37996 USA\\
 Email: \{hdas, ifahad, stushar1, hasib, gbuchan2, dscott57, garose, sai\}@utk.edu
 }}

\maketitle

\begin{abstract}

In this paper, we introduce a novel device architecture that merges memristive devices with light-sensing surfaces, for energy-efficient motion recognition at the edge. Our light-sensing surface captures motion data through in-sensor computation. This data is then processed using a memristive system equipped with a HfO2-based synaptic device, coupled with a winner-take-all (WTA) circuit, tailored for low-power motion classification tasks. We validate our end-to-end system using four distinct human hand gestures—left-to-right, right-to-left, bottom-to-top, and top-to-bottom movements—to assess energy efficiency and classification robustness. Our experiments show that the system requires an average of only \SI{4.17}{\nano\joule} for taking our processed analog signal and mapping weights onto our memristive system and \SI{0.952}{\nano\joule} for testing per movement class, achieving 97.22\% accuracy even under 5\% noise interference. A key advantage of our proposed architecture is its low energy requirement, enabling the integration of energy-harvesting solutions such as solar power for sustainable autonomous operation. Additionally, our approach enhances data privacy by processing data locally, reducing the need for external data transmission and storage.

\end{abstract}

\begin{IEEEkeywords}
Sensor, Memristor, ReRAM, Synapse, NeuroProcessor
\end{IEEEkeywords}


\section{Introduction}

For decades, smart environments have been heralded as transformative, enhancing understanding of context, actions, and activities in various settings such as public areas, offices, homes\cite{smarthome}, workshops, factories, and healthcare centers \cite{Giannetsos2011Nov}. The realization of this vision hinges on the development of autonomous optical sensing technologies, such as light-sensing surfaces, which surpass traditional camera-based methods and von Neumann architectures in energy efficiency, scalability, and privacy preservation. Breakthroughs in these light-sensing surfaces\cite{Zhang2020Sep,Ma2019Aug} have enabled the integration of sensing and computation into everyday objects, enabling applications from gesture-based inputs to non-contact activity detection. Yet, efficiently processing the voluminous data from these sensors while maintaining low power consumption and latency remains a challenge.

A promising approach to mitigate these challenges is in-sensor computation \cite{Zhou2020Nov}, which entails processing sensory data in the analog domain before it undergoes analog-to-digital conversion (ADC) and digital computation. This method compresses high-dimensional data from extensive sensor arrays into manageable forms, reducing redundancy \cite{Kagawa2004Oct,Hsu2019} and addressing privacy concerns by avoiding the transmission of raw data. Such recent advances in sensor computation have enabled computational light-sensing surfaces \cite{Zhang2022Jan} that process raw data into essential vision features such as position, orientation, direction, speed, and identification within the analog domain.

However, despite advances in computational light sensing surfaces, energy demands exist for wireless data transmission in intelligent motion detection systems, necessitating frequent battery changes. Integrating memristive synaptic devices with computational light sensors offers a promising solution for additional energy savings. This may allow for the classification of motion sensor data directly on the device, avoiding the need for power-intensive wireless transmission of raw sensor data.

Therefore, in this work, we propose integrating computational light-sensing surfaces with a memristive system and characterize its performance for power and motion recognition tasks. To the best of our knowledge, this work is the first instance of integrating computational light-sensing surfaces with a memristive system. 

\begin{figure}[t]
    \centering
    \includegraphics[width=1\linewidth]{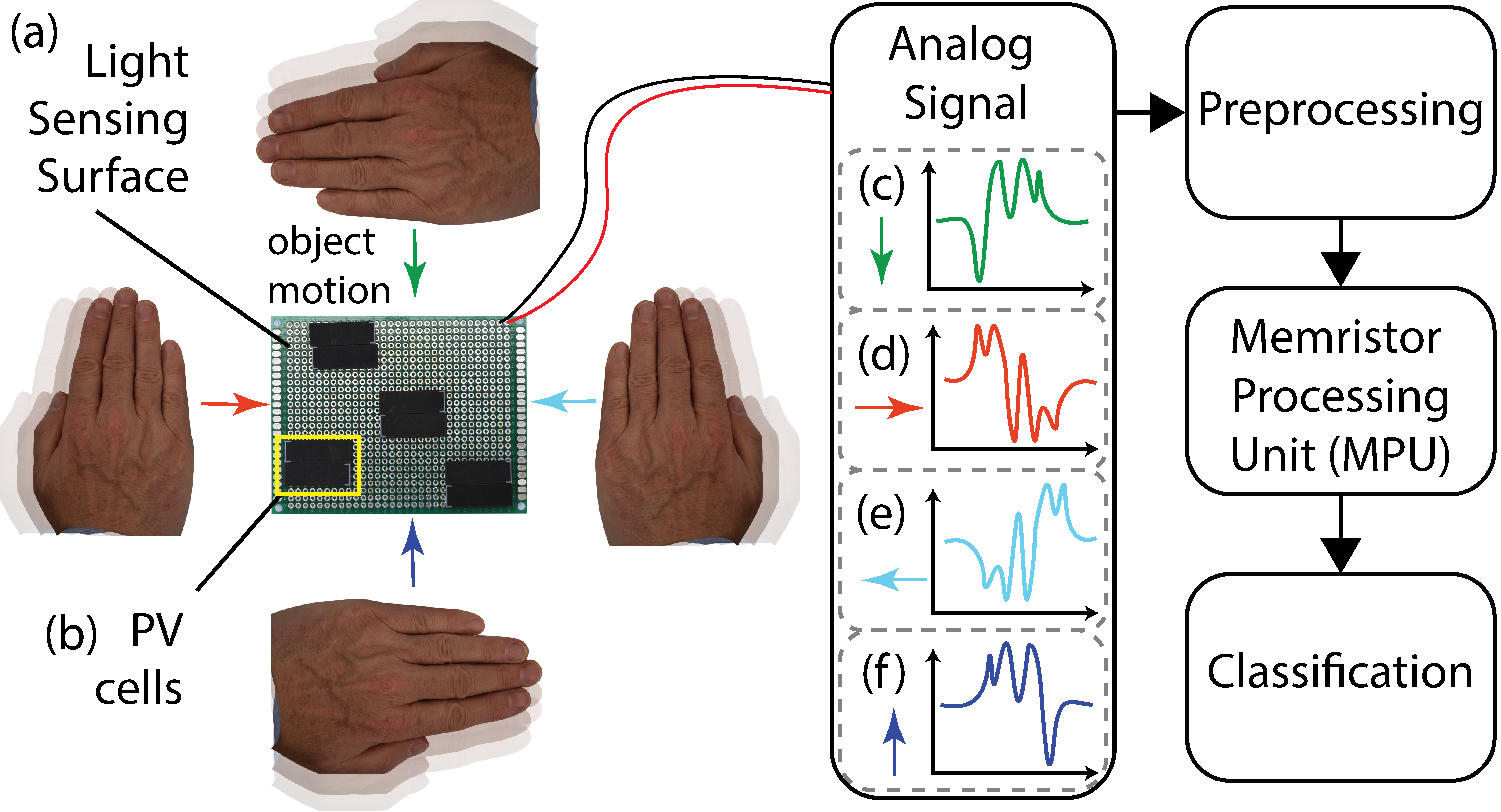}
    \caption{Shows a computational light sensing surface (a) with spatially distributed photovoltaic cells (b) producing a motion-incited analog output for four motions: (c) top to bottom motion (d) left to right motion (e) right to left motion (f) bottom to top motion. The data is then pre-processed and fed to a memristive system for classification and prediction}
    \label{fig:end2end}
    \vspace{-0.5cm}
\end{figure}

We develop our computational light-sensing surfaces utilizing a heterogeneous array of photovoltaic (PV) cells Fig. \ref{fig:end2end}. When an object moves on top of our sensing surface, we achieve in-sensor analog computation by combining the weighted output of PV cells, producing time-resolved differential voltage measurement data. The data is processed by our memeristive system, which we term as Memristive Processing Unit (MPU). Our MPU is designed utilizing \ce{HfO2}-based synaptic devices \cite{das_RFAM} for an energy-efficient design. We use a simple synaptic crossbar capable of in-sensor computation as a dot-product engine (DPE) \cite{das_JETSCAS, das_TCASI}. Once data is processed by the crossbar, we use a winner-take-all (WTA) circuitry as a simple ranking circuit based on the maximum current from the columns (from the synaptic crossbar), which can be utilized for motion classification tasks with a low power budget.  

Consequently, the MPU is employed not only to train the synaptic crossbar but also to test incoming sensory data in real-time, all within a constrained energy budget. To evaluate the robustness of our design, we introduce uniformly random noises to impact the column current of the system.

The key contributions of this paper are as follows.  
\begin{enumerate}
    \item An end-to-end memristive system is proposed for analog signal mapping and testing object motion data. 
    \item Current-controlled synaptic elements are used to construct a reliable and low-power memristive system. 
    \item An evaluation of the proposed system for object motion recognition with real-world distinct motion classes
    \item An empirical validation of classification robustness by testing against various random noises injected in the column current. 
\end{enumerate}

Overall, our system paves the way for a novel approach to energy-efficient motion recognition at the edge.






\subsection{Background on Light Sensing Surfaces}
Historically, light-sensitive surfaces have been created using a network of PV cells and photodiodes, which function as sensors to detect object movement. Objects moving across these surfaces block light to each cell, altering the photocurrent and voltage characteristics. Specifically, PV cells in light exhibit three critical properties: photocurrent ($I_{PD}$), dark current ($I_{D}$, also known as leakage current), and diode capacitance ($C_{D}$). The overall output current ($I_{OUT}$) merges $I_{PD}$ and $I_{D}$, and is converted to a voltage ($V_{PD}$) using a load resistance ($R_{L}$).

In detecting object motion via light sensing, previous systems have used PV cells in two primary modes: a) Photoconductive mode and b) Photovoltaic mode. In photoconductive mode, an external reverse bias voltage aligns $V_{PD}$ directly with light intensity, while reducing diode capacitance $C_{D}$, improving response time. Thus, changes in $V_{PD}$ due to light blockage by objects can be detected. This approach is widely utilized, as in Ma et al.'s SolarGest \cite{ma2019solargest}, which uses PV cells for hand motion detection. Though highly sensitive, it requires external power.

In contrast, the photovoltaic mode functions independently of external power. It relies on the reduced power generation by a photodiode ($V_{PD}$) when light is partially obstructed by an object. Monitoring this change in the photodiode's output power allows for detecting obstructions by near-field objects. This method is effectively used in finger motion recognition, as shown by Li et al. \cite{li2018self} in their self-powered system, and in ambient light sensing surfaces for activity detection by Zhang et al. \cite{zhang2020optosense}.

While effective for gathering raw data on object motion, these techniques face challenges, particularly in data transfer to memory for processing, which aligns with the von Neumann architecture. This transfer can lead to power and latency issues. To address this, our design (Fig. \ref{fig:end2end}(a)), inspired by Zhang et al.'s proposal \cite{zhang2022flexible} to process analog data in-sensor within light-sensing surfaces, enabling these surfaces to be computational.


\subsection{Sensing Principle}

\begin{figure}[t]
    \centering
    \includegraphics[width=0.9\linewidth]{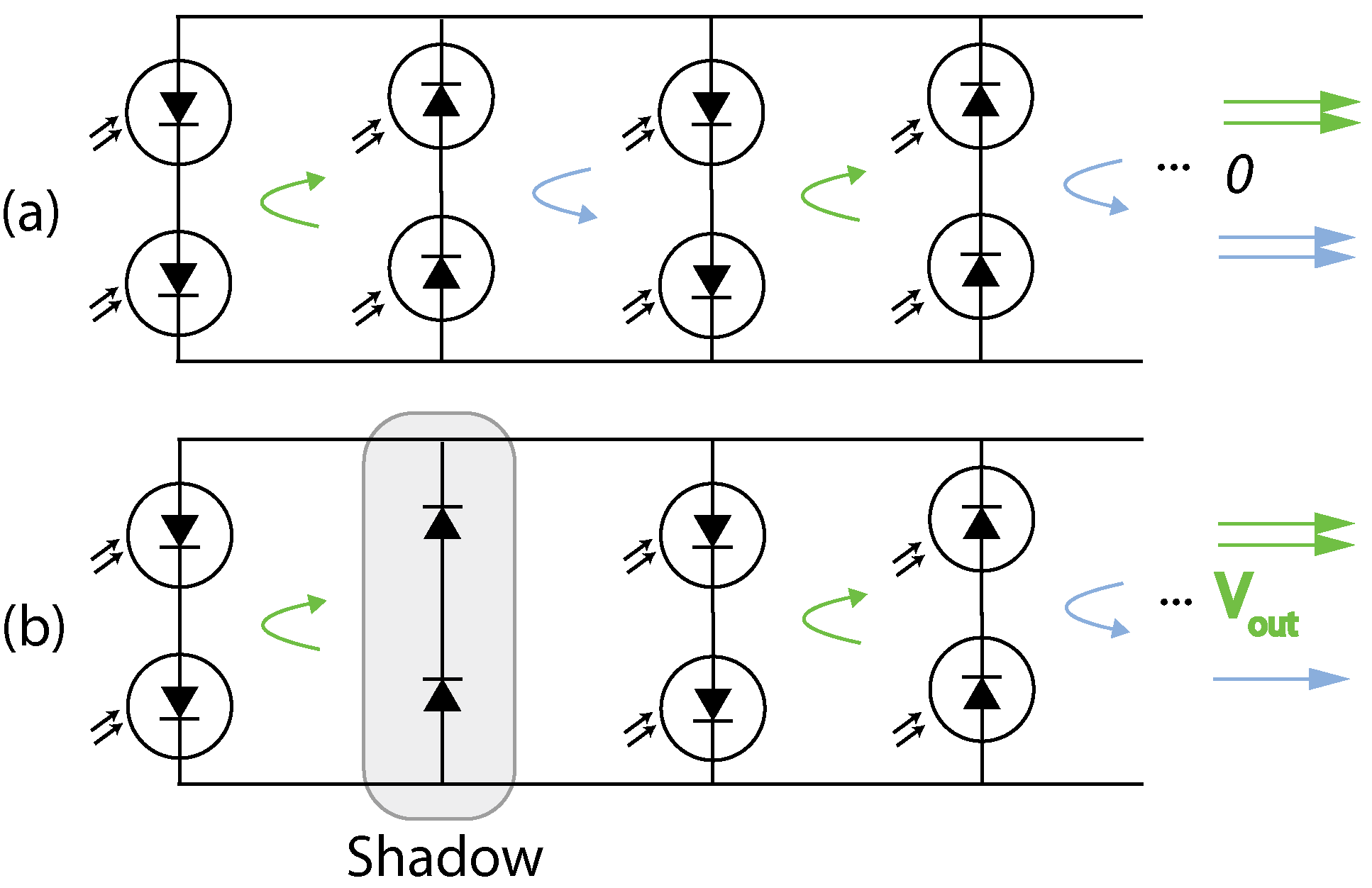}
    \caption{(a) PV cells spatially separated, connected in (serial/parallel) with $V_{out}$ being equal to zero. (b) Variable output due to PV cell occlusion.}
    \label{fig:photodiodeinteraction}
     \vspace{-0.5cm}
\end{figure}

To design a computational light sensing surface, we use a series of interconnected PV cells (Fig. \ref{fig:photodiodeinteraction}), which produce direct current in the presence of light. When PV cells are connected with opposite polarity, the net voltage across the output terminals becomes zero, as shown in Fig. \ref{fig:photodiodeinteraction}(a). 

This configuration allows us to exploit a significant property of the PV cells: their ability to produce varying electrical outputs in response to different light intensities. This variation forms the basis of our motion recognition sensors. When the PV cells, spaced apart strategically, are subjected to varying light conditions due to an object's motion — which causes some cells to be in shadow (Fig. \ref{fig:photodiodeinteraction}(b)) while others remain illuminated — each cell reacts differently based on the light intensity it receives. This leads to a unique pattern of electrical responses across the cells. The overall output voltage ($V_{out}$) from the array of PV cells is then determined according to Kirchhoff's Voltage Law, calculated as the algebraic sum of the individual voltages generated by each cell, which is reflective of the object's motion.

\begin{equation} \label{eq:KVL}
   \text{ $V_{out}$} = V_1 + V_2 + .... + V_n =\sum_{i=1}^{n} V_n = 0 
\end{equation}

\textbf{In-sensor Processing:} Importantly, since the voltage levels are directly associated with the combined voltage output of the photodiodes, influenced dynamically by the motion of objects, our sensors are capable of performing weighted linear combination tasks within the analog domain, generating time-resolved data. Contrasting with traditional sensor systems, where complexity is $\mathcal{O}(\textrm{M})$ -- ``M'' being the number of operations required for digitizing and processing each photodiode's output – our design achieves a reduced complexity of $\mathcal{O}(1)$. This reduced complexity is due to our method of simultaneously acquiring and processing data directly in the analog domain, eliminating the need for separate digitization and processing steps for each photodiode. Consequently, our approach allows for scalable design of different surfaces and shapes with increased photodiode configurations without an increase in system complexity.

\subsection{Computational Light Sensing Surface Design}
Using the principle discussed above, we designed a computational light sensing surface using PV cells (KXOB25-05X3F-TR) that are spatially distributed on the surface of a perfboard as shown in Fig. \ref{fig:end2end}(a). In our design, 4 pairs of identical cells are connected in parallel, and half are in opposite polarity. This pattern can be modeled as a 2D matrix designed to invoke distinctive signal sequences in response to 4 object motion directions: top to bottom (TB), bottom to top (BT), left to right (LR), right to left (RL) gradually blocking, or unblocking sensor PV cells. 

This is analogous to an edge detection kernel, which `convolves' with the signal produced by object motions. In our work, the proposed sensor consists of 4 pairs of spatially distributed PV cells that can work like a $4\times4$ asymmetric matrix: 

\begin{equation*}
    \begin{bmatrix}
        0 & -1 & 0 & 0 \\
        0 & 0 & -1 & 0 \\
        1 & 0 & 0 & 0 \\
        0 & 0 & 0 & 1 \\
    \end{bmatrix}
\end{equation*}

When an object moving in proximity to the top surface of the sensor blocks incident light on an odd number of PV cells, the net voltage at the output terminal can experience three scenarios due to the blocking and unblocking of PV cells such as positive (`$1$'), zero (`$0$') and negative(`$-1$'). When an odd no. of cells is blocked, it produces a positive or negative voltage based on the relative polarity of the connection and becomes zero for an even number. As the object keeps moving in a certain direction, it produces unique sequences such as $[0, 1, 0, -1]$. The size of the sensor board and PV cells can be scaled according to the application requirement.

\subsection{Data Acquisition}

Using the constructed computational light sensing surface, we acquired data for four object motion directions: TB, BT, LR, and RL. The unique signal response of these four different motion directions can be seen in Fig. \ref{fig:end2end} (c), (d), (e), and (f). 


We used a digital oscilloscope (SIGLENT SDS 1202X-E) to obtain analog signals produced by the interaction of motions with the light-sensing surface. The analog signals are acquired at \SI{1}{\mega\hertz} using the oscilloscope to verify the distinguishable patterns of object motion from different directions. The data is collected in the laboratory at ambient light illumination (\SI{500}{\lux} to \SI{1200}{\lux}). The amplitude of the analog signals varies from \SI{-1.14}{\volt} to \SI{1.22}{\volt} at that light level. Object motion data were collected \SI{2}{\centi\meter} above the top surface of the sensor, and the motion speed is $0.5$ to \SI{0.8}{\m\per\s}. 


\textbf{Pre-Processing}: After obtaining the analog sensor response data, we sampled it at \SI{30}{\milli\s} time intervals selecting eight samples. These samples were then utilized to map weights into our memristive system for motion recognition tasks.

\section{Proposed Hardware}

We now review the components of the proposed hardware, including the Memristive Processing Unit (MPU) used for mapping weights and testing analog data from our computational light-sensing surfaces. 

\subsection{Memristive Synapse}
\begin{figure}[t]
\centering
\includegraphics[width=2.8in, height=1.5in]{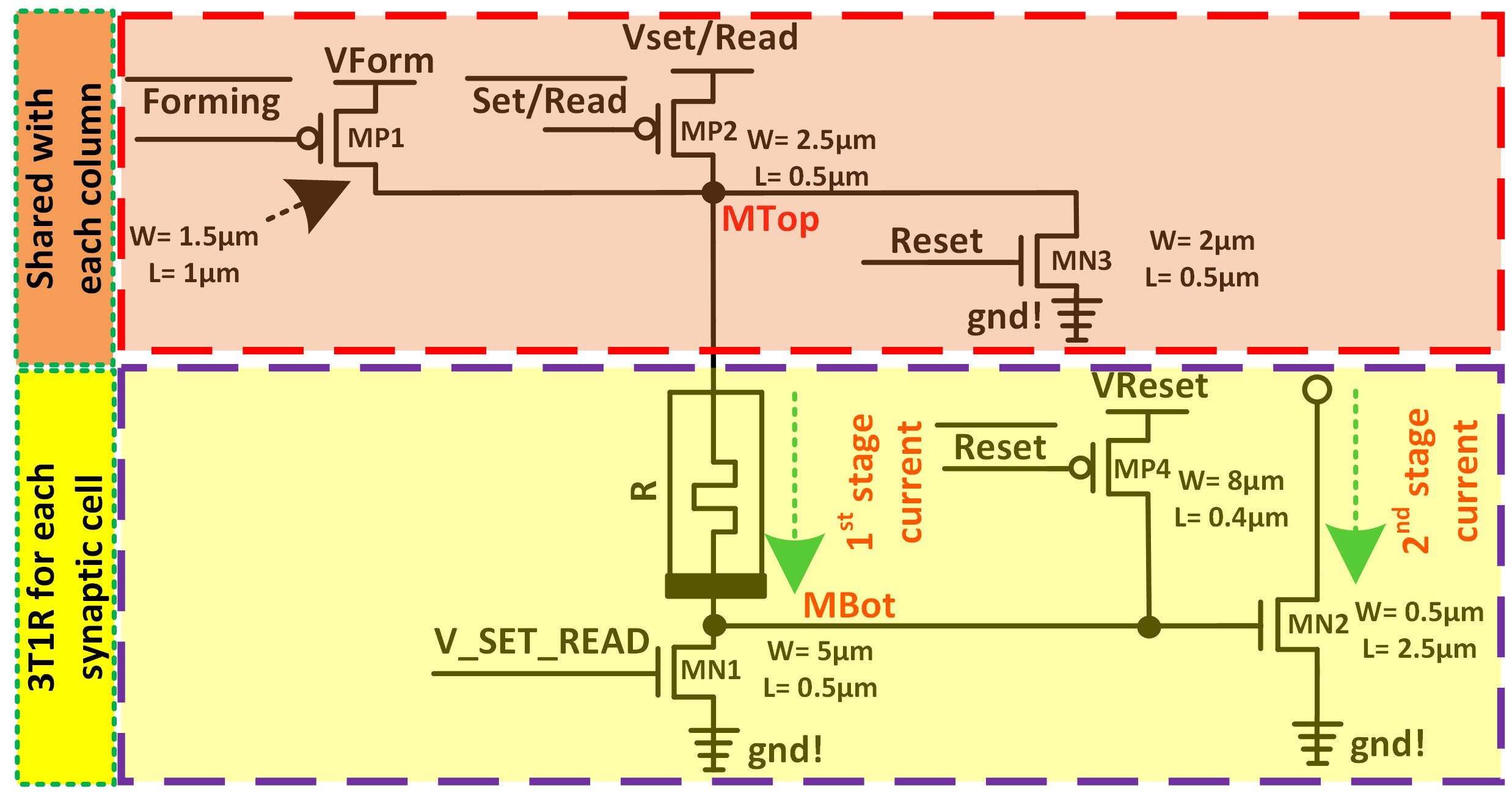}
\caption{CMOS-based memristive synapse. Each synapse contains three transistors, a memristive device, and a control block to perform $FORM$, $RESET$, $SET$, and $READ$ operations. The control block is connected at the top of the memristor which is denoted as MTop.}
\vspace{-0.5cm}
\label{fig:synapse}
\end{figure}

A \ce{HfO2}-based memristive devices are used to construct the synapses. Verilog-A model is utilized to simulate the synaptic circuits on Cadence Specter with \SI{65}{\nano\meter} CMOS technology from IBM\cite{das_RFAM}. Fig. \ref{fig:synapse} shows the CMOS and memristive device-based synapse. At first, the synaptic devices will be at their higher resistance state (HRS). A one-time $FORM$ operation creates a filament from the top to the bottom electrode. After a successful $FORM$ operation, the resistance level of the synapse will be a few \SI{}{\kilo\ohm}. Then a $RESET$ operation is initiated to set the resistance level from a few \SI{}{\kilo\ohm} to a few hundred \SI{}{\kilo\ohm}. Then the synapse will be programmed or $SET$ from \SI{5}{\kilo\ohm} to \SI{20}{\kilo\ohm} with a $SET$ voltage from \SI{0.7}{\volt} to \SI{1.2}{\volt} at $V\_SET\_READ$. All operations such as $FORM$, $RESET$, and $SET$ are operated with a pair of CMOS. Now, the synapse is ready for $READ$ operation. Here, $READ$ is a two-stage operation. As seen in Fig. \ref{fig:synapse}, at the first stage, MP2 and MN1 are utilized to initiate the $READ$ operation with a \SI{0.6}{\volt} at $V\_SET\_READ$. The first stage current will be converted to a voltage to drive MN2. Finally, the second stage current will be sensed from the drain of the MN2. Different $READ$ currents will be generated for different resistance levels.    

\begin{figure}[t]
\centering
\includegraphics[width=0.9\linewidth]{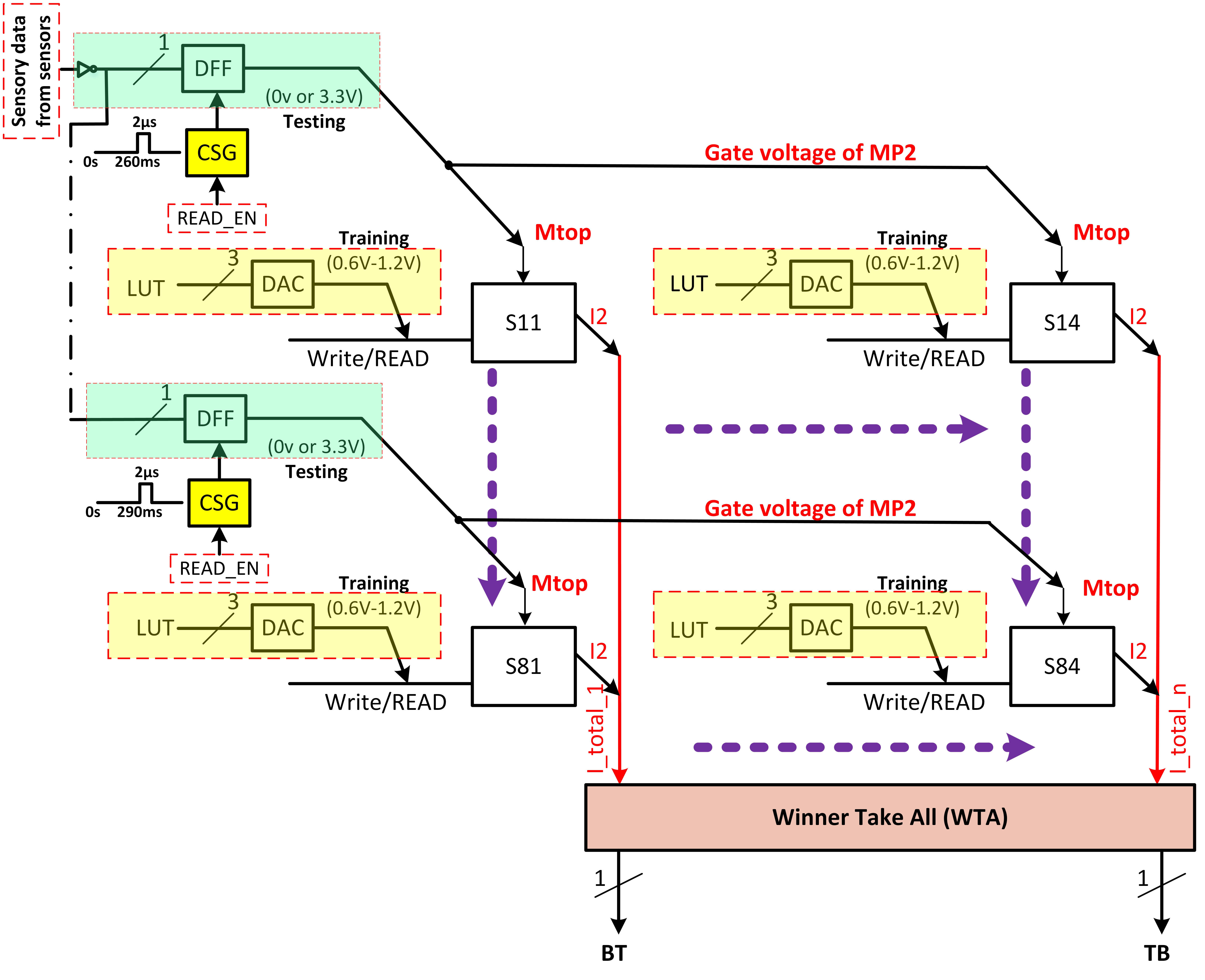}
\caption{Synaptic crossbar with peripherals to train the system and test the sensor data in runtime. 4-bit DACs are utilized to train the system according to sampled data from the light-sensing surface signals. DFFs and customized signal generators are used to capture the data from sensors and $READ$ the columns of the crossbar simultaneously.}
\label{fig:arc}
\end{figure}

\subsection{Synaptic Crossbar with Peripherals}
This work uses $4\times8$ synaptic devices to construct a crossbar or DPE. Fig. \ref{fig:arc} shows the synaptic crossbar, which is configured as S11 to S84. Each column has 8 synaptic elements, where Mtops are sharing the control circuity row-wise. Initially, all the synaptic components will be formed and reset. Here, the $V\_SET\_READ$ signal is denoted as $Write\ /READ$. A digital-to-analog converter (DAC) with a 4-bit resolution is used to program memristive synapses with different weight levels. D-flip-flops (DFF) are used to READ the weights column-wise. At the same time, column-wise cumulative currents will be sensed as $I\_total\_1$, $I\_total\_2$, and so on. Finally, the summed current from different columns will be forwarded to a WTA circuitry to determine the class based on the maximum current.       

\section{Analog Signal Mapping Approach}

Table \ref{tab:sampling} shows the pre-processed data collected from the computational light sensing surface, and it is sampled with \SI{30}{\milli\s} clock cycle. The sampling is started at \SI{230}{\milli\s} and sampled for \SI{2}{\micro\s}. For each object motion class, data will be sampled off-chip, and 8 random samples will be selected to represent a specific movement. For example, when the object moves from bottom to top (BT) on a computational light sensing surface, a distinctive voltage signal that correlates with the passage of time will be collected. Then, the signal will be sampled for 8 pulse widths. Then, 8 DACs are utilized to map the eight synaptic elements of a column with 8 sampled pulses, which are different in amplitudes. The average energy for 4-bit DACs is about \SI{282.56}{\pico\joule} for each column, where the analog voltage can be varied from \SI{0.6}{\volt} to \SI{1.2}{\volt}. 

\begin{table}[t]
    \centering
    \caption{Sampled Data from Computational Light Sensing Surface}
    \label{tab:sampling}
    \resizebox{\columnwidth}{!}{
    \begin{tabular}{@{}|c|c|c|c|c|c|@{}}
    \toprule
    \textbf{\makecell{Time \\Interval}} & \textbf{\makecell{Sampled \\for}} & \textbf{\makecell{Amplitude \\of BT$^*$}} & \textbf{\makecell{Amplitude \\of LR$^*$}} & \textbf{\makecell{Amplitude \\of RL$^*$}} & \textbf{\makecell{Amplitude \\of TB$^*$}} \\ \midrule
    \SI{260}{\milli\s} &  \SI{2}{\micro\s} & \SI{1.16}{\volt}  & \SI{-0.60}{\volt} & \SI{0.28}{\volt}  & \SI{-0.24}{\volt}   \\ \midrule
    \SI{290}{\milli\s} &  \SI{2}{\micro\s} & \SI{1.10}{\volt}  & \SI{0.22}{\volt}  & \SI{0.48}{\volt}  & \SI{-1.10}{\volt}    \\ \midrule
    \SI{320}{\milli\s} &  \SI{2}{\micro\s} & \SI{0.30}{\volt}  & \SI{0.78}{\volt}  & \SI{0.74}{\volt}  & \SI{-1.08}{\volt}  \\ \midrule
    \SI{350}{\milli\s} &  \SI{2}{\micro\s} & \SI{-0.58}{\volt} & \SI{0.44}{\volt}  & \SI{0.88}{\volt}  & \SI{-0.88}{\volt}    \\ \midrule
    \SI{380}{\milli\s} &  \SI{2}{\micro\s} & \SI{-1.04}{\volt} & \SI{0.68}{\volt}  & \SI{0.70}{\volt}  & \SI{-0.50}{\volt}   \\ \midrule
    \SI{410}{\milli\s} &  \SI{2}{\micro\s} & \SI{-0.96}{\volt} & \SI{0.62}{\volt}  & \SI{0.52}{\volt}  & \SI{0}{\volt}    \\ \midrule
    \SI{440}{\milli\s} &  \SI{2}{\micro\s} & \SI{-0.36}{\volt} & \SI{-0.02}{\volt} & \SI{0.78}{\volt}  & \SI{0.34}{\volt}   \\ \midrule
    \SI{470}{\milli\s} &  \SI{2}{\micro\s} & \SI{-0.16}{\volt} & \SI{0.06}{\volt}  & \SI{-0.02}{\volt} & \SI{1.2}{\volt}    \\ \midrule
    \end{tabular}
    }
    \footnotesize{$^*$ BT = Bottom to Top, LR = Left to Right, RL = Right to Left, TB = Top to Bottom}\\
\end{table}

\begin{table}[t]
    \centering
    \caption{Weight or resistance level after mapping analog data}
    \label{tab:training}
    \resizebox{\columnwidth}{!}{
    \begin{tabular}{@{}|c|c|c|c|c|c|@{}}
    \toprule
    \textbf{\makecell{Time \\Interval}} & \textbf{\makecell{Sampled \\for}} & \textbf{\makecell{Weight \\of BT}} & \textbf{\makecell{Weight \\of LR}} & \textbf{\makecell{Weight \\of RL}} & \textbf{\makecell{Weight \\of TB}} \\ \midrule
    \SI{260}{\milli\s} &  \SI{2}{\micro\s} & \SI{5.10}{\kilo\ohm}  & \SI{96.29}{\kilo\ohm} & \SI{96.29}{\kilo\ohm} & \SI{96.29}{\kilo\ohm}  \\ \midrule
    \SI{290}{\milli\s} &  \SI{2}{\micro\s} & \SI{5.15}{\kilo\ohm}  & \SI{96.29}{\kilo\ohm} & \SI{96.29}{\kilo\ohm} & \SI{96.29}{\kilo\ohm}  \\ \midrule
    \SI{320}{\milli\s} &  \SI{2}{\micro\s} & \SI{96.29}{\kilo\ohm} & \SI{27.73}{\kilo\ohm} & \SI{44.69}{\kilo\ohm} & \SI{96.29}{\kilo\ohm} \\ \midrule
    \SI{350}{\milli\s} &  \SI{2}{\micro\s} & \SI{96.29}{\kilo\ohm} & \SI{96.29}{\kilo\ohm} & \SI{11.30}{\kilo\ohm} & \SI{96.29}{\kilo\ohm}  \\ \midrule
    \SI{380}{\milli\s} &  \SI{2}{\micro\s} & \SI{96.29}{\kilo\ohm} & \SI{96.34}{\kilo\ohm} & \SI{78.35}{\kilo\ohm} & \SI{96.29}{\kilo\ohm}  \\ \midrule
    \SI{410}{\milli\s} &  \SI{2}{\micro\s} & \SI{96.29}{\kilo\ohm} & \SI{96.32}{\kilo\ohm} & \SI{96.29}{\kilo\ohm} & \SI{96.29}{\kilo\ohm}  \\ \midrule
    \SI{440}{\milli\s} &  \SI{2}{\micro\s} & \SI{96.29}{\kilo\ohm} & \SI{96.29}{\kilo\ohm} & \SI{27.73}{\kilo\ohm} & \SI{96.29}{\kilo\ohm}  \\ \midrule
    \SI{470}{\milli\s} &  \SI{2}{\micro\s} & \SI{96.29}{\kilo\ohm} & \SI{96.29}{\kilo\ohm} & \SI{96.29}{\kilo\ohm} & \SI{5.11}{\kilo\ohm}  \\ \midrule
    \end{tabular}
    }
    \footnotesize{$^*$ Before the weight updates all the synaptic devices are $RESET$ at \SI{\sim100}{\kilo\ohm}}\\
    \footnotesize{$^*$ All the weight updates are based on the Cadence Spectre simulation with \SI{65}{\nano\meter} CMOS technology from IBM 10LPe}\\    
\end{table}

Following the mapping of the first column, another signal from left to right (LR) is collected from the computational light sensing surface and sampled with the same time intervals (\SI{2}{\micro\s}) as the first signal (BT). After sampling, signal data from LR is mapped to column two using DACs. In this manner, all the columns of our designed crossbar are mapped with weights for different analog motion signals. In addition, if all the memristive synapses are mapped from \SI{1.2}{\volt} to \SI{0.6}{\volt}, the average energy consumption of a column is about \SI{3.89}{\nano\joule} for \SI{2}{\micro\s} clock cycle. The updated weights of the mapped columns are illustrated in Table \ref{tab:training}. We now outline the approach for testing our system, which processes incoming data from a computational light-sensing surface using our proposed architecture.                    

\section{Testing Approach}

\begin{table}[htb]
    \centering
    \caption{$READ$ current from synapses during testing}
    \label{tab:testing}
    \resizebox{\columnwidth}{!}{
    \begin{tabular}{@{}|c|c|c|c|c|c|@{}}
    \toprule
    \textbf{\makecell{Time \\Interval}}  & \textbf{\makecell{Sampled \\for}}   & \textbf{Column:1}          & \textbf{Column:2}        & \textbf{Column:3}        & \textbf{Column:4}           \\ \midrule
    \SI{260}{\milli\s}      &  \SI{2}{\micro\s}      & \SI{5.827}{\micro\ampere}  & \SI{3.564}{\micro\ampere} & \SI{3.564}{\micro\ampere} & \SI{3.564}{\micro\ampere}    \\ \midrule
    \SI{290}{\milli\s}      &  \SI{2}{\micro\s}      & \SI{5.825}{\micro\ampere}  & \SI{3.564}{\micro\ampere} & \SI{3.564}{\micro\ampere} & \SI{3.564}{\micro\ampere}    \\ \midrule
    \SI{320}{\milli\s}      &  \SI{2}{\micro\s}      & \SI{3.564}{\micro\ampere}  & \SI{5.221}{\micro\ampere} & \SI{4.776}{\micro\ampere} & \SI{3.564}{\micro\ampere}    \\ \midrule
    \SI{350}{\milli\s}      &  \SI{2}{\micro\s}      & \SI{3.564}{\micro\ampere}  & \SI{4.452}{\micro\ampere} & \SI{5.658}{\micro\ampere} & \SI{3.564}{\micro\ampere}    \\ \midrule
    \SI{380}{\milli\s}      &  \SI{2}{\micro\s}      & \SI{3.564}{\micro\ampere}  & \SI{4.241}{\micro\ampere} & \SI{3.958}{\micro\ampere} & \SI{3.564}{\micro\ampere}    \\ \midrule
    \SI{410}{\milli\s}      &  \SI{2}{\micro\s}      & \SI{3.564}{\micro\ampere}  & \SI{3.567}{\micro\ampere} & \SI{3.564}{\micro\ampere} & \SI{3.564}{\micro\ampere}    \\ \midrule
    \SI{440}{\milli\s}      &  \SI{2}{\micro\s}      & \SI{3.564}{\micro\ampere}  & \SI{3.569}{\micro\ampere} & \SI{5.211}{\micro\ampere} & \SI{3.564}{\micro\ampere}    \\ \midrule
    \SI{470}{\milli\s}      &  \SI{2}{\micro\s}      & \SI{3.564}{\micro\ampere}  & \SI{3.586}{\micro\ampere} & \SI{3.564}{\micro\ampere} & \SI{5.828}{\micro\ampere}    \\ \midrule
    \ \makecell{If incoming \\signal  = BT}  &  \SI{2}{\micro\s}  & \textcolor{red}{\SI{11.96}{\micro\ampere}}   & \SI{7.428}{\micro\ampere} & \SI{7.428}{\micro\ampere} & \SI{7.428}{\micro\ampere}    \\ \midrule
    \ \makecell{If incoming \\signal  = LR}  &  \SI{2}{\micro\s}  & \SI{10.95}{\micro\ampere}  & \textcolor{red}{\SI{13.28}{\micro\ampere}} & \SI{12.55}{\micro\ampere} & \SI{10.95}{\micro\ampere}    \\ \midrule
    \ \makecell{If incoming \\signal  = RL}  &  \SI{2}{\micro\s}  & \SI{23.75}{\micro\ampere}  & \SI{24.72}{\micro\ampere} & \textcolor{red}{\SI{26.83}{\micro\ampere}} & \SI{21.49}{\micro\ampere}    \\ \midrule
     \ \makecell{If incoming \\signal  = TB}  &  \SI{2}{\micro\s}  & \SI{3.92}{\micro\ampere}  & \SI{3.92}{\micro\ampere} & \SI{3.92}{\micro\ampere} & \textcolor{red}{\SI{6.19}{\micro\ampere}}    \\ \midrule
    
    \end{tabular}
    }
\footnotesize{$^*$ $READ$ activation depends on the incoming signals from the sensors.}\\
\footnotesize{$^*$ All the columns are representing individual synaptic currents.}\\
    
\end{table}


Our testing approach supports a real-time operation based on the data coming from the computational light-sensing surface. As seen in Fig. \ref{fig:arc}, the data from the computational light sensing surface is stored with DFFs using a customized signal generator (CSG) by capturing the first sample in the first DFF, the second sample in the second DFF, and so on. After the signal is sampled 8 times and values are stored in 8 DFFs, the stored values of the DFFs are propagated for all the rows and turns on the $READ$ circuitry from the MTop. DFFs will make the output either \SI{0}{\volt} or \SI{3.3}{\volt} to the gate of the $MP2$ for a $READ$ operation. If the amplitude of the sampled signal is below \SI{0.45}{\volt}, then it is treated as \SI{3.3}{\volt} with an inverter, otherwise the output signal of the DFFs will be \SI{0}{\volt}. The sampling circuit consumes \SI{3.24}{\pico\joule} energy on average per class of object motion (e.g., BT, LR, etc.) from the computational light sensing surface. If all the synaptic cells are storing values from \SI{5.10}{\kilo\ohm} to \SI{96.34}{\kilo\ohm} in a column, then the average $READ$ energy is \SI{90.16}{\pico\joule} for a column. 

When all the columns are $READ$ together, each column produces a cumulative current based on the input $READ$ signal. According to Table \ref{tab:testing}, only the column matched with sensor data will produce the maximum amount of current among all the columns. For example, if the incoming sensor data is representing BT signal, then only the $1^{st}$ and $2^{nd}$ DFF will provide \SI{0}{} to the gate of the $READ$ MOSFET. In addition, the rest of the DFF will sample \SI{3.3}{\volt} from $3^{rd}$ to $8^{th}$ and keep the synaptic device in standby mode. Stand-by synaptic devices consume about \SI{51}{\nano\ampere} current during a $READ$ operation. A class or movement of the object will be determined by a WTA based on the maximum current from the columns. According to Table \ref{tab:testing}, the average classification operation consumes \SI{116.38}{\pico\J}, \SI{140.7}{\pico\J}, \SI{200.98}{\pico\J}, and \SI{103.84}{\pico\J} when the incoming signals are BT, LR, RL, and TB respectively. The average energy consumption with WTA is about \SI{140.48}{\pico\joule} per classification. The total energy consumption with the CSG with DFF, crossbar, and WTA is about \SI{0.952}{\nano\joule} per classification.                   


\section{Performance Evaluation and Comparison with Prior Work}

We evaluated the robustness of our system by analyzing the effects of noise on analog data acquired for four distinct signal types: BT, LR, RL, and TB. The WTA circuit is only effective when the maximum column current does not overlap with currents from other columns. The overlap can occur when noise reduces the maximum column current and increases the current of other columns, even though the maximum current was distinguishable without the noise. Therefore, we perform noise analysis by adding 1\%, 3\%, and 5\% Gaussian noise to the column current in such a way that the maximum column current decreases by 1\%, 3\%, and 5\% due to noise, while the current of other columns increases by 1\%, 3\%, and 5\% due to noise. We generated 81 scenarios for each signal type (BT, LR, RL, and TB) by combining three variations each of noise levels, their effects on maximum and other column currents, and their interaction with the other signal types. 

We ran our simulations in Cadence. We observed in the second test case, when the incoming signal is LR, in 9 out of 81 scenarios, the WTA mechanism fails to distinguish classes. This occurs only when the column with the maximum current decreases by 5\% due to Gaussian noise. Overall, we observed 97.22\% classes can be detected accurately with 5\% noise. However, for the other test cases, the presence of 1\%, 3\%, and 5\% noise does not affect the outcome. 

When juxtaposed with prior work, as shown in Table \ref{tab:comparison}, our proposed design demonstrates significant energy efficiency. It consumes merely \SI{0.952}{\nano\joule} per classification task, a calculation that includes all peripherals. This assessment was conducted to handle four classes of analog signals at a \SI{33}{\hertz} sampling rate. In contrast, the HfO\textsubscript{x}-based neuromorphic system for motion detection, as detailed in \cite{wang2021neuromorphic} and utilizing CMOS-based image sensors, exhibits an energy consumption of \SI{87.5}{\nano\joule}, which is about $92\times$ higher than our method. Another comparative study, which employed DYNAP-SE for EMG classification of three gestures, reported an energy expenditure of \SI{19.91}{\nano\joule}, nearly $21\times$ greater than our proposed design with \SI{180}{\nano\meter} process.

\begin{table}[t]
    \centering
    \caption{Comparison with prior work}
    \label{tab:comparison}
    \resizebox{\columnwidth}{!}{
    \begin{tabular}{@{}|c|c|c|c|c|c|@{}}
    \toprule
    \textbf{Reference} & \textbf{Architecture} & \textbf{Sensor} & \textbf{\makecell{\# of\\ class}} & \textbf{Energy} & \textbf{\makecell{Sample\\ frequency}} \\ \midrule
    AIS'20 \cite{wang2021neuromorphic}               & HfO\textsubscript{x} & CMOS-based & 4   & \SI{87.5}{\nano\joule} & -   \\ \midrule
    AICAS'20\cite{ma2020neuromorphic} &  DYNAP-SE & - &   3   & \SI{19.91}{\nano\joule} &-   \\ \midrule
    This work &  \makecell{Synaptic (\ce{HfO2})\\ crossbar} & PD  & 4 & \SI{0.952}{\nano\joule} & \SI{33}{\hertz} \\ \midrule
    \end{tabular}
     }
    \footnotesize{$^*$ Energy numbers are based on a classification task}\\ 
    \end{table}

\section{Conclusions and Future Work}

We propose an energy-efficient motion detection system integrating memristive devices with light-sensing surfaces. Our approach achieves impressive energy savings compared to previous systems. Furthermore, the current-controlled memristive synapses in our design demonstrate robust performance in classification tasks, even in the presence of external noise affecting the column current. Remarkably, the system maintains an accuracy of approximately 97.22\% in object motion classification tasks, even with the addition of up to 5\% uniform random noise. Looking ahead, the synaptic device design introduced in this work lays a foundational framework for enhancing accuracy, generalizability, and robustness by exploring various learning techniques within the neuromorphic domain. This includes methodologies like spike-timing-dependent plasticity \cite{Nishith_MWSCAS, STDP1, STDP3, STDP4, IOP}, homeostatic plasticity \cite{Homeo}, and reservoir computing \cite{RC, ACSSC_2023, das2024enhanced}, all promising avenues for enhancing activity detection using neuromorphic computing. In addition, more complex motion detection test cases where sensors and objects are both in motion with various light intensities can be considered for future evaluation.    

\section*{Acknowledgments}
This material is based partly on research sponsored by the Air Force Research Laboratory under agreement FA8750-21-1-1018. The U.S. Government is authorized to reproduce and distribute reprints for Governmental purposes, notwithstanding any copyright notation thereon. The views and conclusions contained herein are those of the authors and should not be interpreted as necessarily representing the official policies or endorsements, either expressed or implied, of the Air Force Research Laboratory or the U.S. Government. 

\bibliographystyle{IEEEtran}
\bibliography{refs.bib}

\end{document}